\documentclass{elsart}

\usepackage{graphicx}
\usepackage{fancyhdr}
\usepackage{setspace}

\begin{document}

\runauthor{Carjan, Talou, Serot}

\begin{singlespace}









\begin{frontmatter}

\title{Emission of Scission Neutrons in the Sudden Approximation}

\author[CENBG]{N.~Carjan}
\author[CAD,LANL]{P.~Talou}\footnote{Corresponding author: patrick.talou@cea.fr}
\author[CAD]{O.~Serot}

\address[CENBG]{Universit\'e Bordeaux 1; CNRS/IN2P3; \\ Centre d'Etudes Nucl\'eaires de Bordeaux Gradignan, UMR 5797, \\ Chemin du Solarium, BP 120, 33175 Gradignan, France}

\address[CAD]{CEA-Cadarache, DEN/DER/SPRC/LEPh, B\^at. 230, 13108 St-Paul-lez-Durance, France}

\address[LANL]{Nuclear Physics Group, Theoretical Division, Los Alamos National Laboratory, \\ Los Alamos, NM 87545, U.S.A.}

\begin{abstract}
At a certain finite neck radius during the descent of a fissioning nucleus from the saddle to the scission point, the attractive nuclear forces can no more withstand the repulsive Coulomb forces producing the neck rupture and the sudden absorption of the neck stubs by the fragments. At that moment, the neutrons, although still characterized by their pre-scission wave functions, find themselves in the newly created potential of their interaction with the separated fragments. Their wave functions become wave packets with components in the continuum. The probability to populate such states gives evidently the emission probability of neutrons at scission. In this way, we have studied scission neutrons for the fissioning nucleus $^{236}$U, using two-dimensional realistic nuclear shapes. Both the emission probability and the distribution of the emission points relative to the fission fragments strongly depend on the quantum numbers of the pre-scission state from which the neutron is emitted. In particular it was found that states with $\Omega \pi$ = 1/2+ dominate the emission. Depending on the assumed pre- and post-scission configurations and on the emission-barrier height, 30 to 50\% of the total scission neutrons are emitted from 1/2+ states. Their emission points are concentrated in the region between the newly separated fragments. The upper limit for the total number of neutrons per scission event is predicted to lie between 0.16 and 1.73 (depending on the computational assumptions).
\end{abstract}

\end{frontmatter}

{\bf keywords:} Nuclear fission,  scission neutrons,  sudden approximation.

\end{singlespace}

\newpage

\section{Introduction}

When a cold nucleus undergoes fission, the coupling between the collective and intrinsic motions can either be adiabatic (the nucleons remain in the lowest energy levels) or diabatic (the nucleons are promoted to excited states). This is mainly governed by the velocity of the collective motion relative to the Fermi velocity.

At the beginning of the descent from saddle to scission, the Coulomb repulsion is compensated to a large extent by the nuclear attraction. The coupling between the collective and intrinsic motions is therefore weak (quasi-adiabatic): the collective motion is relatively slow and only few nucleons are excited with increasing deformation.

Close to scission however this quasi-equilibrium breaks down. The neck between the fragments ruptures and it is quickly absorbed by the fragments. The coupling is then diabatic.

In the present work, we are making the extreme assumption that the fission motion is adiabatic until the neck ruptures at finite radius and that this rupture is followed by the sudden absorption of the neck protuberances. At that moment a transition between two quite different nuclear configurations occurs. The corresponding large difference in the potential energy of deformation is made available to the internal degrees of freedom, that is, mainly to the nucleons but also to preformed clusters. \\
This sudden release was proposed as an emission mechanism by Halpern~\cite{Halpern:1965a} without being quantitatively developed. The only attempt so far to carry out calculations using the sudden approximation was done later using a simplistic one-dimensional model~\cite{Serot:2000a} for the emission of scission alpha particles. A gradual potential change and its effect on the energy transferred to the nucleons was studied by Fuller~\cite{Fuller:1962a} and by Boneh and Fraenkel~\cite{Boneh:1974a}.

In the present paper we have generalized the formalism from Ref.~\cite{Serot:2000a} to two dimensions in order to include realistic nuclear shapes, and have applied it to the emission of neutrons during $^{235}$U ($n_{th}$,f) reaction.

Emission of scission nucleons has always been an intriguing subject. Based on energetic considerations they should be the most probable among the light particles that accompany the fission process. In particular they should outnumber the alpha particles. Experimentally however the emission rate for scission protons represents only few percent of that for scission alphas~\cite{Wagemans:1991a} while the existing results on scission neutrons are very contradictory~\cite{Wagemans:1991a,Bowman:1962a}. Understanding this paradox is a central piece of the nuclear fission puzzle.

For scission neutrons, there is a clear need for more precise experimental data. Nevertheless, in a recent paper, Kornilov~\cite{Kornilov:2001a} has re-analyzed three independent experiments on $^{252}$Cf(sf) with a more realistic model for the description of the neutron evaporation from the fragments. In all these experiments, the neutron energy and angular distributions relative to the fragment direction were measured. Kornilov showed that a good agreement exist between these experiments, and that a neutron surplus of (30 $\pm$ 5)\% exists at about 90$^\circ$ relative to the direction of the moving fragments. These neutrons do not originate from fully accelerated fragments, and would represent about 10\% of the total fission neutrons yield. A similar conclusion has been drawn from angular distribution of neutrons measured by Franklyn~\cite{Franklyn:1978a} for the $^{235}$U(n$_{th}$,f) reaction. This angular distribution could not be reproduced if scission neutrons are neglected but could be very well fitted with a scission neutron contribution to the total neutron yield of 20\%.

It is also worth noting that using a systematics of ternary charged particle yields established by Val'ski~\cite{Valski:2003a}, scission neutron yields can be extrapolated for various fissionning nuclei. These scission neutron yields are found to be (0.18 $\pm$ 0.04)/fission for $^{252}$Cf(sf) and (0.55 $\pm$ 0.09)/fission for $^{235}$U(n$_{th}$,f), representing about 4.8\% and 22.7\% of the total fission neutrons yields, respectively. A recent review of experimental data on the scission neutron characteristics can be found in~\cite{Petrov:2005a}.

In the next section, we will show how to calculate primary fragments excitation energies, scission neutrons multiplicities and spatial distributions of the emission points starting from the single-neutron wave functions at the two nuclear configurations between which the sudden emission is supposed to occur. Section 3 contains our predictions for the low energy fission of U-236 obtained for independent and pairing correlated neutrons, for two sets of initial and final configurations, and for two energy cuts separating bound from unbound states. In section 4, we will give a brief summary of the main results.

\section{The formalism of the sudden approximation}

The single-particle wave functions for an axially-symmetric fissioning nucleus have the general form
\begin{eqnarray} \label{eq:psi}
\Psi \left( \rho,z,\phi \right) = u(\rho,z)e^{i \left(\Omega-\frac{1}{2}\right)\phi}|\uparrow>+d(\rho,z)e^{i \left(\Omega+\frac{1}{2}\right)\phi}|\downarrow>,
\end{eqnarray}
where $u(\rho,z)$ and $d(\rho,z)$ contain the spatial dependence of the two components, spin up and down respectively. $\Omega$ is the projection of the total angular momentum along the symmetry axis and is a good quantum number. Since we are dealing with symmetric fission, the parity $\pi$ is also a constant of motion.

If the scission is characterized by a sudden change of nuclear deformation, an eigenstate of the "just before scission" hamiltonian will be distributed over the eigenstates of the "immediately after scission" hamiltonian:
\begin{eqnarray}
|\Psi^i> = \sum_{all states}{a_{if}|\Psi^f>},
\end{eqnarray}
where
\begin{eqnarray}\label{eq:coef}
a_{if}=< \Psi^f | \Psi^i >=2\pi\int \int \left( u^iu^f+d^id^f\right)\rho d\rho dz.
\end{eqnarray}
One can notice that only $|\Psi^f>$ states with the same $(\Omega,\pi)$ values as $|\Psi^i>$ will have non-zero contributions. These states are mainly bound states but contain also a few discrete states in the continuum that will spontaneously decay. Therefore the emission probability of a neutron that had occupied the state $|\Psi^i>$ is
\begin{eqnarray}\label{eq:Pem}
P_{em}^i = \sum_{unbound states}{\left| a_{if}\right|^2} = 1 - \sum_{bound states}{\left| a_{if}\right|^2}.
\end{eqnarray}
In this way we do not need to use the states in the continuum that are less precise in the numerical diagonalization. The $P_{em}^i$'s will be referred as partial probabilities since they only consider one occupied state. The limit between bound and unbound states is the barrier for neutron emission, which is zero if the centrifugal potential is null. At this point it is worth noticing that our criterion for emission is based on energetic considerations, while Halpern's was based on spatial considerations.

Summing over all occupied states one obtains the total number of scission neutrons per fission event
\begin{eqnarray} \label{eq:Nn}
N_n = \sum_i v_i^2P_{em}^i,
\end{eqnarray}
where $v_i^2$ is the ground-state occupation probability of $|\Psi^i>$.

We have also calculated the part of the initial wave function that was emitted
\begin{eqnarray} \label{eq:psi_em_i}
|\Psi_{em}^i> = |\Psi^i> - \sum_{bound states}{a_{if}|\Psi^f>},
\end{eqnarray}
since it gives access to the spatial distribution of the emission points
\begin{eqnarray} \label{eq:total-emission}
S_{em}(\rho,z) = \sum_i{v_i^2 \left| \Psi_{em}^i (\rho,z)\right|^2}.
\end{eqnarray}
Of course, we have
\begin{eqnarray}
\int{S_{em}(\rho,z)\rho d\rho dz} = N_n.
\end{eqnarray}

Finally, the occupation probabilities after the sudden transition are obtained by summing the contributions of all initial states that populate a given final state:
\begin{eqnarray} \label{eq:exproba}
V_f^2 = \sum_i{v_i^2 \left| a_{if} \right|^2}.
\end{eqnarray}
They are of interest since they show the degree of excitation in which the fragments are left:
\begin{eqnarray} \label{eq:exenergy}
E^*_{FF}=\sum_{bound st.}{V_f^2e^f}-\sum_{bound st.}{v^2_fe^f},
\end{eqnarray}
where $e^f$ are the energies of the bound eigentstates of the hamiltonian ``immediately after scission''.

\section{Numerical results}

We have applied the formalism described above to quantitatively study the emission of scission neutrons during the low-energy fission of $^{236}$U. 

\begin{figure}[ht]
\centerline{\rotatebox{-90}{\includegraphics[width=0.85\textwidth]{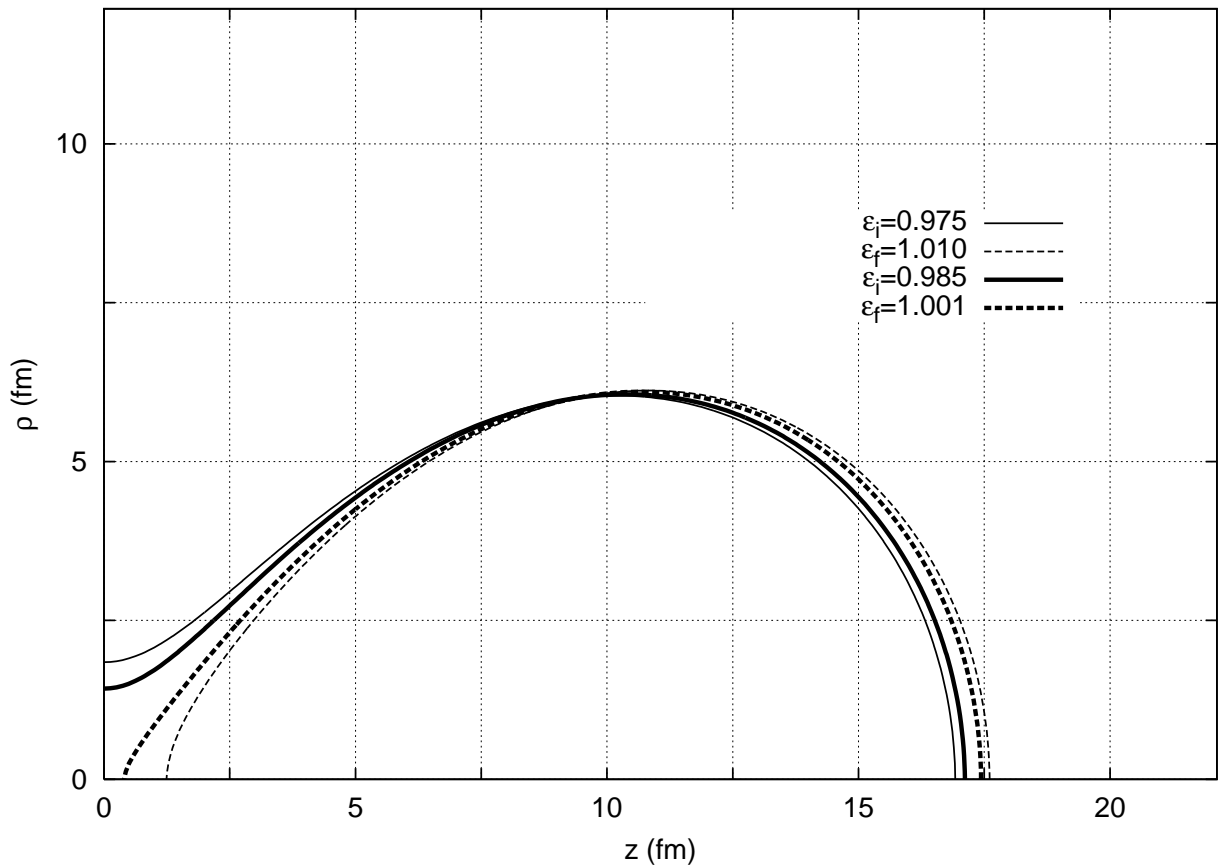}}}
\caption{\label{fig:pot} Equipotentials $V_0/2$ for the two scission configurations studied in this work. The radius of the neck at scission is 1.5~fm for the configuration (0.985,1.001) and 1.9~fm in the case of (0.975,1.010).}
\end{figure}

The nuclear shapes involved were chosen to be Cassini ovals since they require only one deformation parameter $\epsilon$ to describe realistic scission configurations. To see how critical is the choice of the two deformations between which the sudden transition is supposed to occur, we have used two sets of values ($\epsilon_i,\epsilon_f$), 0.975 $\rightarrow$ 1.010 and 0.985 $\rightarrow$ 1.001, describing a longer and a shorter jump respectively ($\epsilon$=1 describes a zero-neck scission shape). An idea of the shapes involved is given (Fig.~\ref{fig:pot}) by the neutron equipotential lines corresponding to $V_0/2$, $V_0$=-40.2 MeV being the depth of the Woods-Saxon interaction. The neck radii at rupture are 1.9~fm and 1.5~fm respectively.

The corresponding hamiltonians have been diagonalized in a deformed oscillator basis and complete sets of eigenfunctions ${|\Psi^i(\epsilon_i)>}$, ${|\Psi^f(\epsilon_f)>}$ and eigenvalues ${e^i(\epsilon_i)}$, ${e^f(\epsilon_f)}$ have been obtained~\cite{Pashkevich:1971a}. From these, only the bound states ($e<0$) enter in the present calculations. It turned out that $^{236}$U has 98 such states at all four considered deformations.

\subsection{Partial emission probabilities}

\begin{figure}[ht]
\centerline{\includegraphics[height=0.9\textheight]{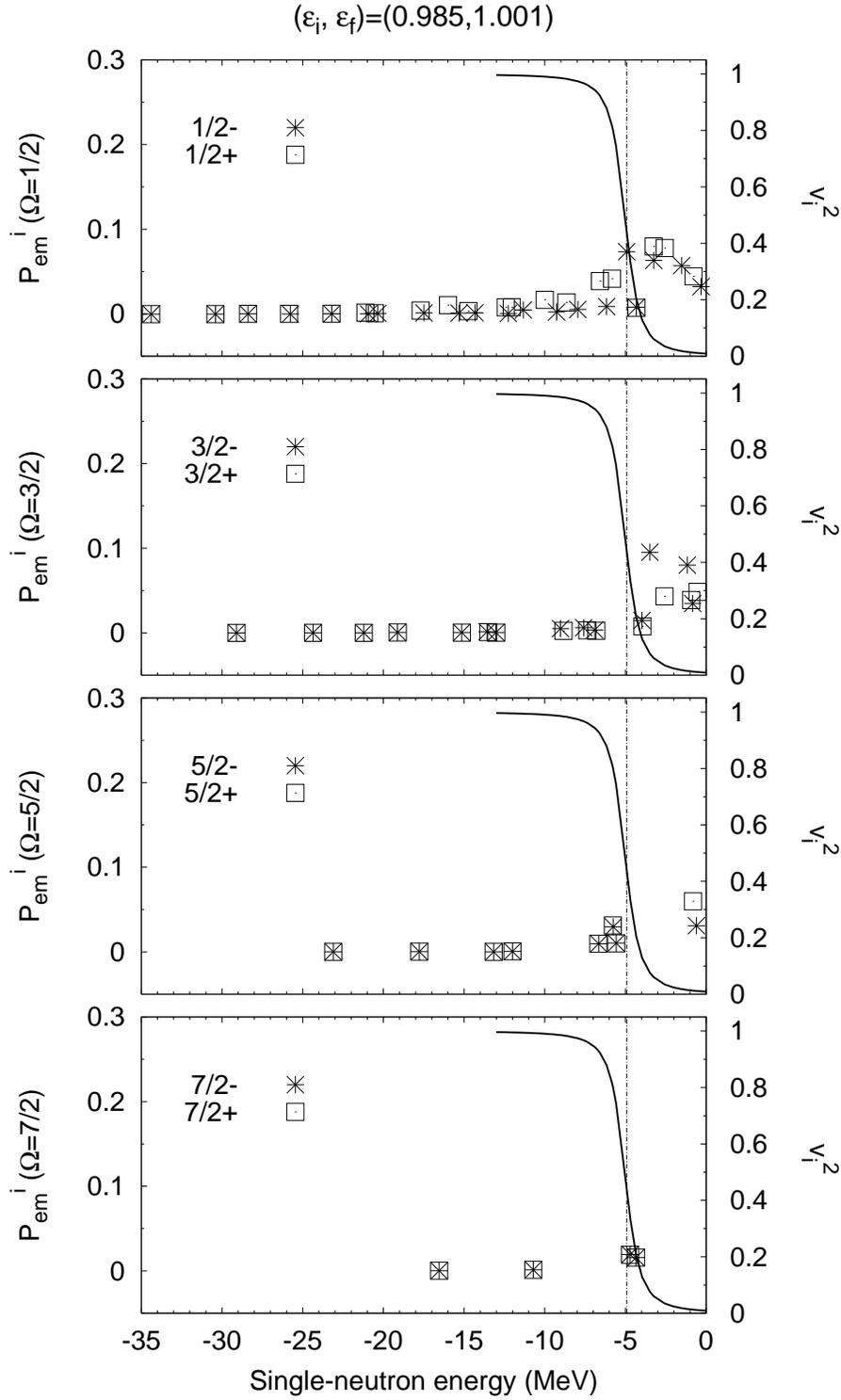}}
\caption{\label{fig:Pem.985} Partial emission probabilities $P_{em}^i$ for the configuration $(\epsilon_i,\epsilon_f)$=(0.985,1.001). The solid line represents the single-particle occupation probabilities in the initial configuration, calculated in the BCS formalism (right scale). The dashed line represents the location of the Fermi energy at -4.926 MeV.}
\end{figure}

\begin{figure}[ht]
\centerline{\includegraphics[height=0.9\textheight]{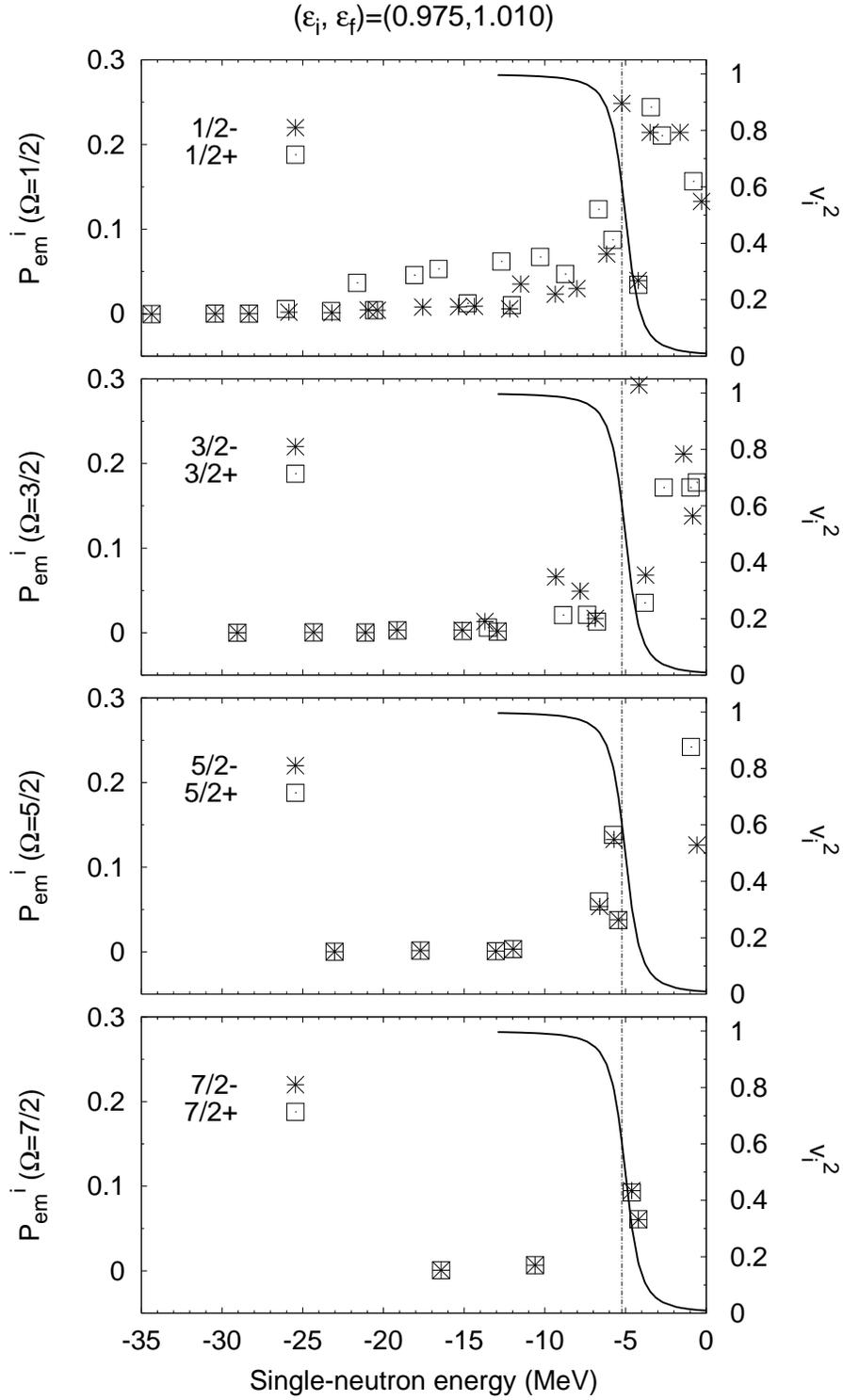}}
\caption{\label{fig:Pem.975} Same as in Fig.~\ref{fig:Pem.985} for the configuration $(\epsilon_i,\epsilon_f)$=(0.975,1.010). The value of the Fermi energy is now -5.229 MeV.}
\end{figure}

For each initial bound state we have calculated the expansion coefficients $a_{if}$, Eq.~(\ref{eq:coef}), corresponding to all final bound states. Using these coefficients, we have calculated the partial emission probabilities, Eq.~(\ref{eq:Pem}). The results are presented in Figures \ref{fig:Pem.985} and \ref{fig:Pem.975} for the transitions 0.985$\rightarrow$1.001 and 0.975$\rightarrow$1.010 respectively. They are separated according to their quantum numbers.

As expected the longer jump produces more scission neutrons. One can also notice that most of these neutrons are emitted from $\Omega=1/2$ states with a preference for $\pi=+$. The variation of $P_{em}^i$ with the energy of the single-neutron state is different for deep lying states than for states around the Fermi energy ($e_F \simeq$-5 MeV). At the beginning (up to about -8 MeV), there is a slow increase; afterwards, the increase becomes steeper and attains a maximum (at about -4 MeV) followed by a moderate decrease. Although this type of variation is present for all $\Omega$-values, the absolute values of the partial emission probabilities are decreasing with increasing $\Omega$.

Over this average trend there are fluctuations due to the structure of the neutron states. To understand these features one has to take a closer look at the neutron wave functions involved.

\begin{figure}[ht]
\centerline{\includegraphics[width=\textwidth]{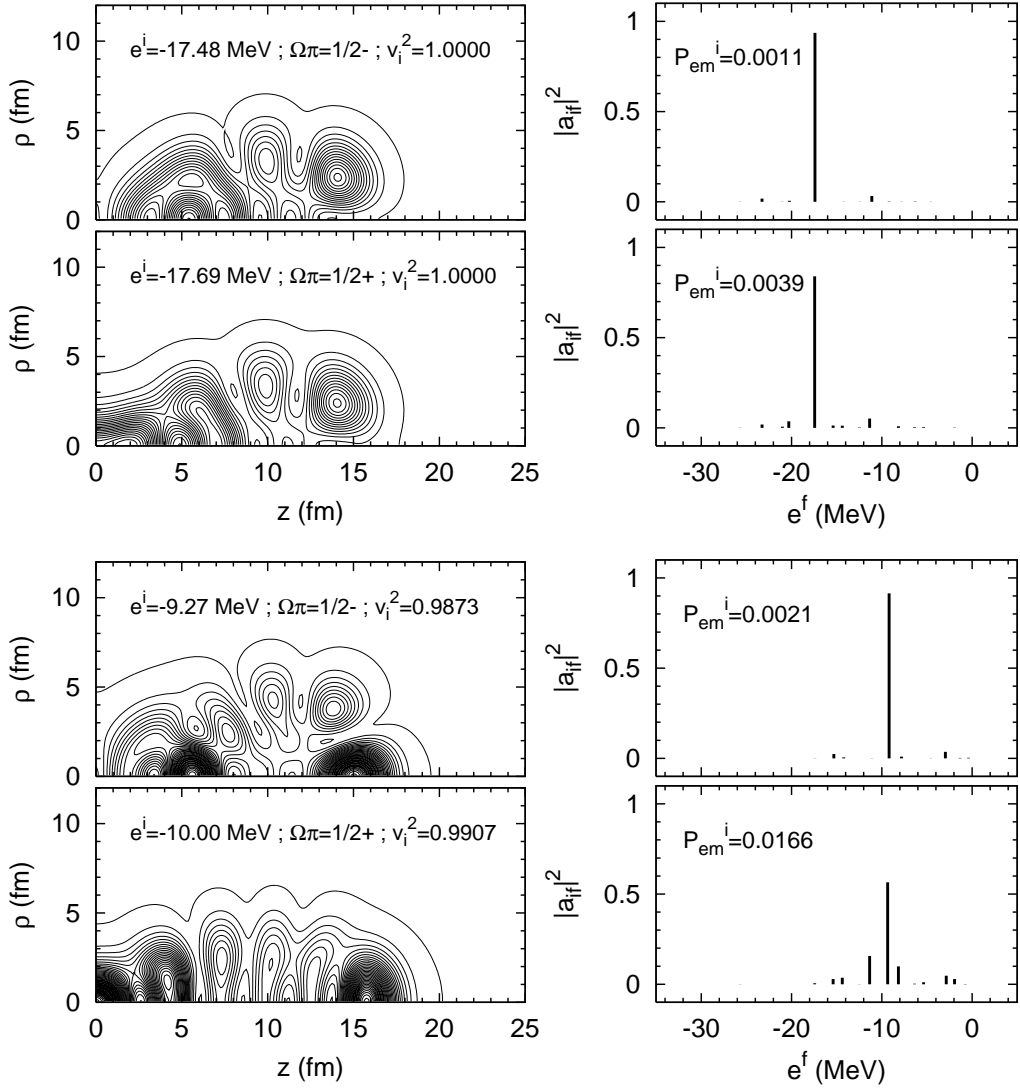}}
\caption{\label{fig:wfs1}Two pairs of low-lying initial states (left) and their expansion coefficients on the final bound states (right) for the configuration (0.985,1.001). Initial states with a high probability of presence in the neck are strongly emitted (large $P_{em}^i$) since they can only expand on final states in the continuum.}
\end{figure}

\begin{figure}[ht]
\centerline{\includegraphics[width=1.2\textwidth]{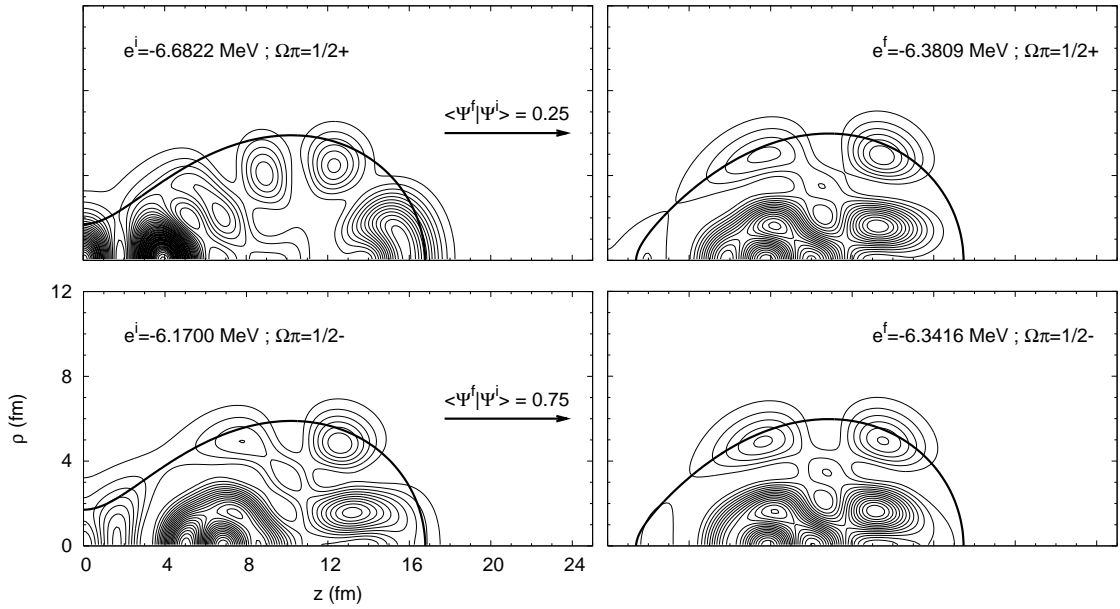}}
\caption{\label{fig:pair} A pair of wave functions at $\epsilon_i=0.975$ with the corresponding adiabatic pair at $\epsilon_f=1.010$. Initial states that are localized in the neck (upper graph) lead to large emission probabilities. On the contrary, states mostly present outside the neck (lower graph) are already in a post-scission configuration, and are much less emitted.}
\end{figure}

Figure~\ref{fig:wfs1} shows two pairs of deep initial states together with their distribution of probabilities over final bound states. A pair is defined by two states that only differ by parity. The sudden transition was supposed to occur between $\epsilon_i=0.985$ and $\epsilon_f=1.001$. A very low partial emission probability is associated with an initial wave function that is essentially distributed only over bound final wave functions. In general, this situation occurs when the initial wave-function resembles very much its adiabatic-partner final wave function. The wave packet in the final configuration is very narrow and the neutron remains essentially bound. Negative-parity initial states can more easily satisfy this condition since they cancel in the origin as do the states of the newly separated fragments. Hence they are already in a post-scission configuration, and this explains why they are less emitted than the corresponding positive-parity states. For a better understanding of this last remark, we have compared in Fig.~\ref{fig:pair} a pair of wave functions at $\epsilon_i$=0.975 with the corresponding adiabatic pair at $\epsilon_f$=1.010.

\begin{figure}[ht]
\centerline{\includegraphics[width=\textwidth]{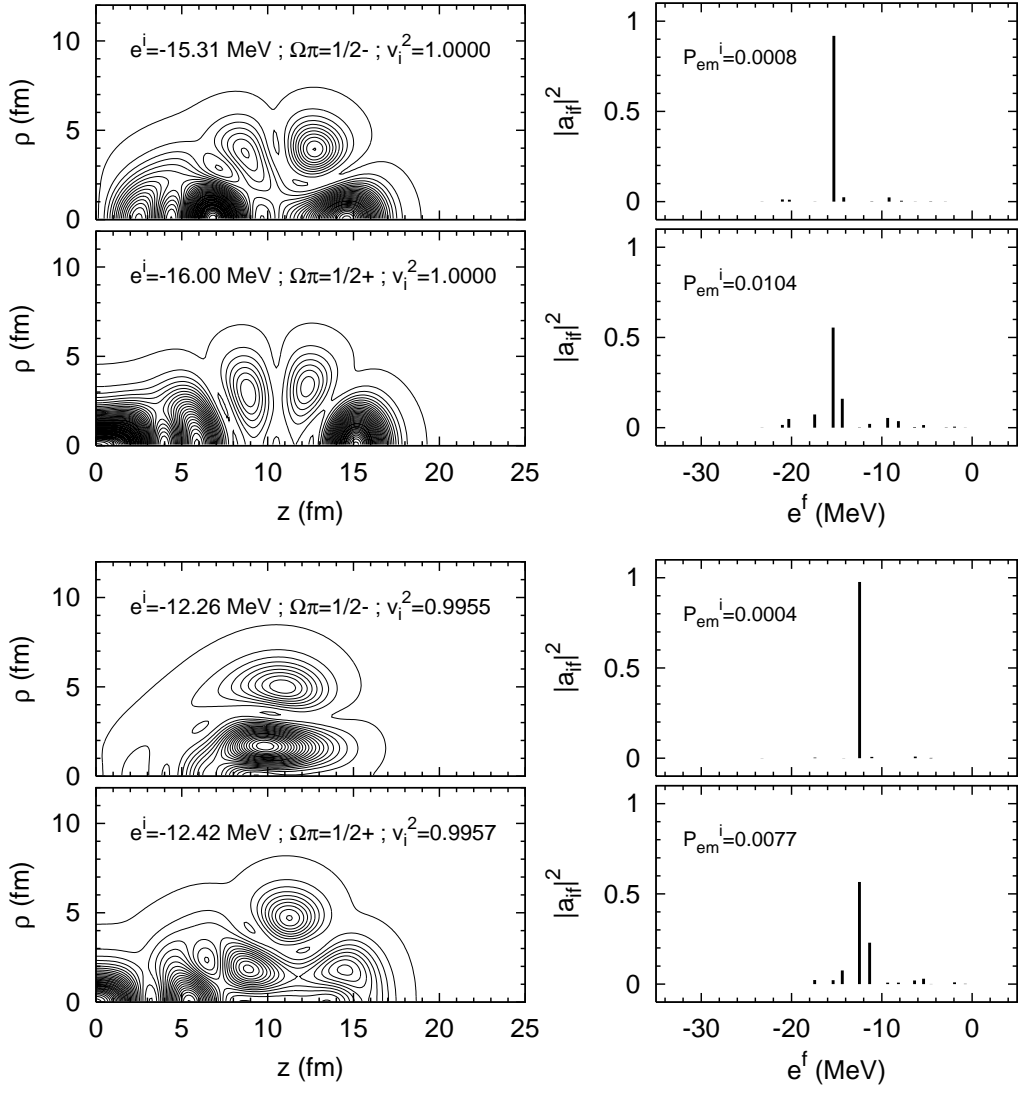}}
\caption{\label{fig:wfs2}Same as Fig.~\ref{fig:wfs1} for two different pairs of initial states. Although the states represented in the upper graph lie lower in energy than the ones in the lower graph, they have a higher emission probability, due to nuclear structure effects.}
\end{figure}

Figure~\ref{fig:wfs2} shows a case in which the nuclear structure effects are strong enough to invert the average trend. The states that are more bound (-15.3110 MeV and -15.9955 MeV) have a higher emission probability than the states that are less bound (-12.2621 MeV and -12.4234 MeV) since, as it can be seen, the latter wave packets are narrower.

\begin{figure}[ht]
\centerline{\includegraphics[height=0.85\textheight]{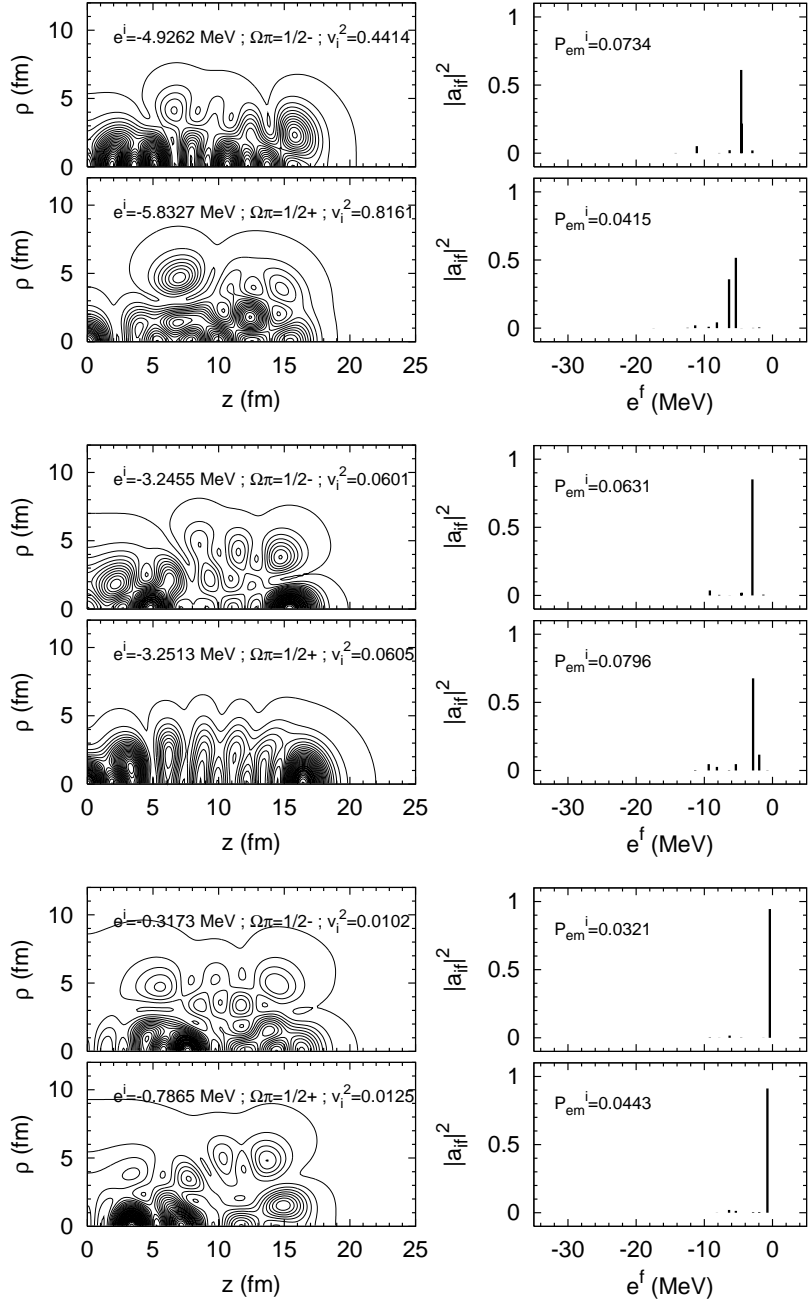}}
\caption{\label{fig:wfs3}Evolution of the emission probability spectrum as a function of the energy of the initial states.}
\end{figure}

Figures~\ref{fig:Pem.985} and~\ref{fig:Pem.975} show that $P_{em}^i (\Omega =1/2)$ exhibits a maximum at about - 3~MeV. When the energy of the initial state increases there are fewer bound states with higher energy left in the final configuration and the distribution of $|a_{if}|^2$ becomes asymmetric, i.e. the higher-energy part of the distribution is cut at zero (see right-hand side of Fig.~\ref{fig:wfs3}). Therefore the wave packet has to include more unbound states and this leads to an increase in the emission probability. However, loosely bound states (-0.3173 MeV and -0.7885 MeV) lie too high to be much perturbed by the sudden increase of the potential in the neck region (see Fig.~\ref{fig:wfs3}). It can be seen that, similar to the strongly bound states (Fig.~\ref{fig:wfs1}) these states are essentially distributed over only one final state and have therefore a smaller partial emission probability. Moreover such states are practically unoccupied ($v_i^2\simeq 0$).

\begin{figure}[ht]
\centerline{\includegraphics[height=0.85\textheight]{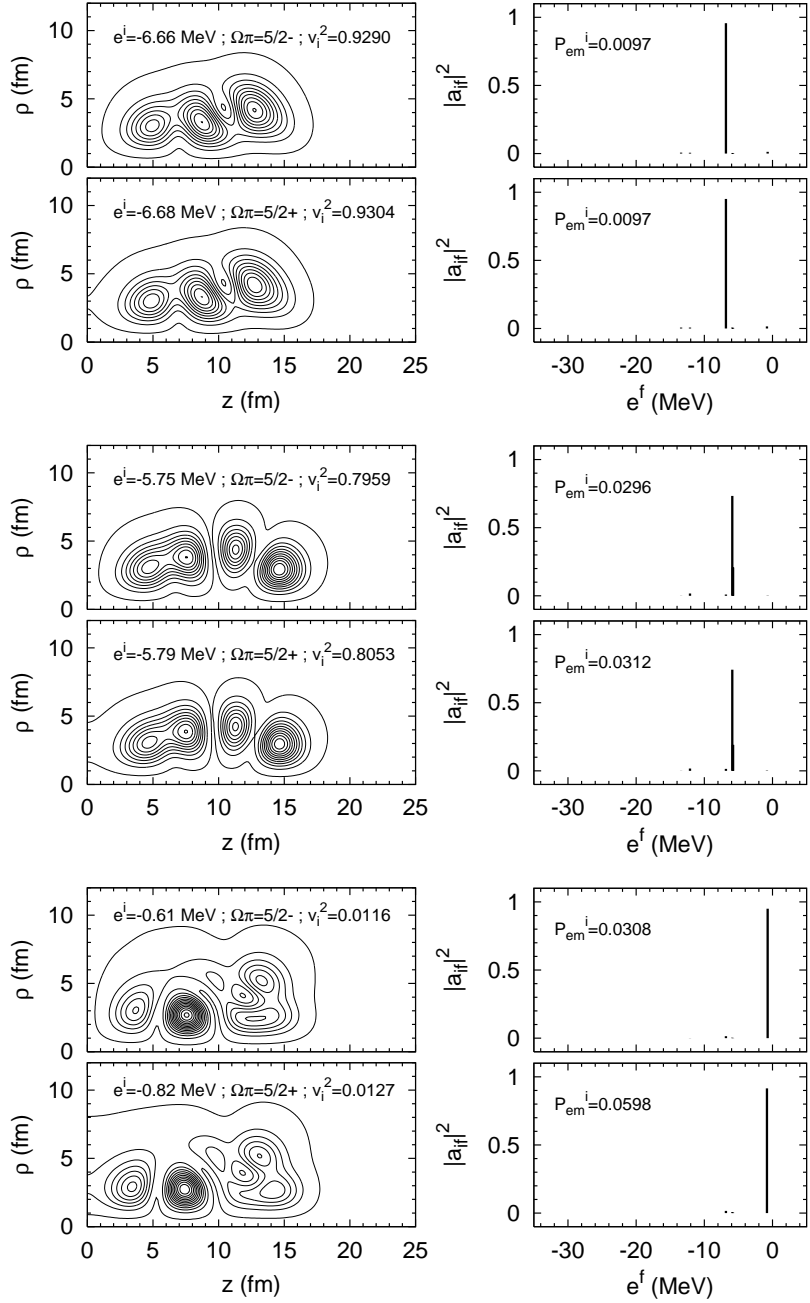}}
\caption{\label{fig:wfs4}States with $\Omega=5/2$ are less emitted than $\Omega=1/2$ states. The presence of the term $(\Omega\pm 1/2)^2/\rho^2$ in the deformed potential pushes the wave functions away from the $z$-axis and strongly hinders $|\psi|^2$ in the neck region.}
\end{figure}

\begin{figure}[ht]
\centerline{\includegraphics[width=\textwidth]{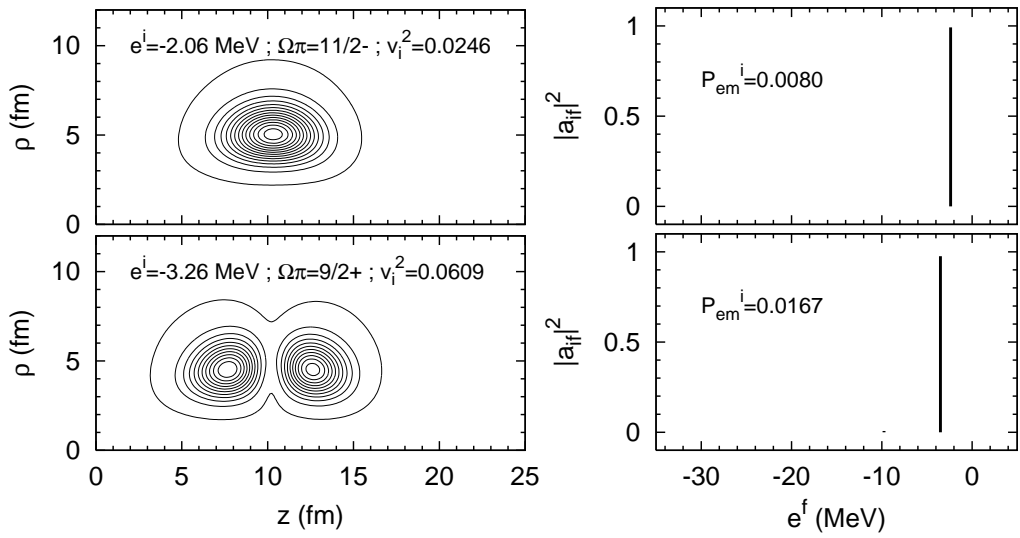}}
\caption{\label{fig:wfs5}States with even higher $\Omega$-values (here, 9/2 and 11/2) cannot be present in the neck region. Therefore, their emission probabilities are negligible.}
\end{figure}

Figure~\ref{fig:wfs4} shows why states with higher $\Omega$-values (5/2) are less emitted. The presence in the deformed potential of a term in $\left( \Omega \pm 1/2\right)^2/\rho^2$ (reminiscent of the centrifugal potential) pushes the neutron away from the z-axis. Therefore their presence in the neck is reduced even for $\pi=+$ states. Of course states with the highest projections of the angular momentum ($\Omega=9/2$ and $\Omega=11/2$) lie very far from the z-axis and they can only be located in the future fragments (see Fig.~\ref{fig:wfs5}). Consequently, their emission probability is negligible.

\begin{figure}[ht]
\centerline{\includegraphics[width=\textwidth]{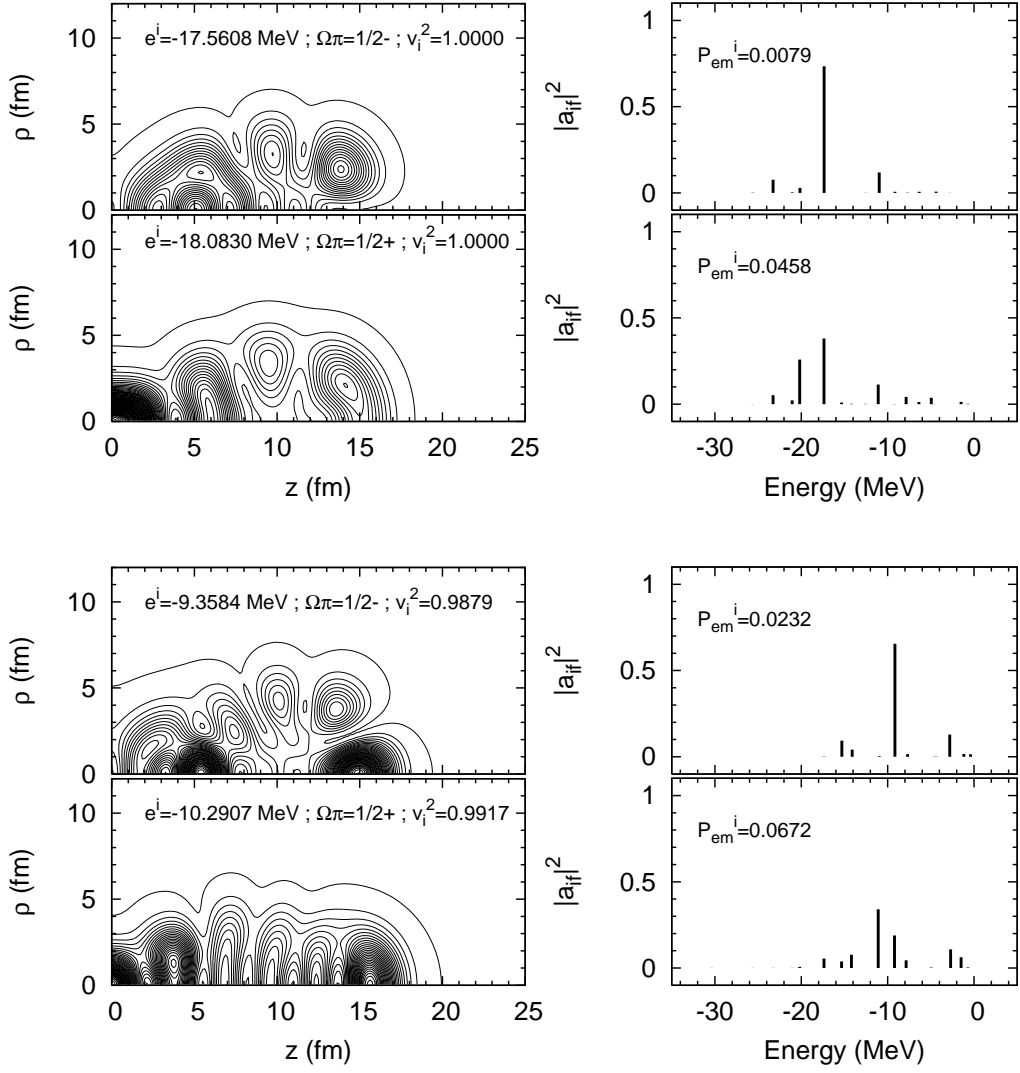}}
\caption{\label{fig:wfs6}The states corresponding to those of Fig.~\ref{fig:wfs1} are represented here for the configuration $(0.975,1.010)$. As expected, the larger difference between the "just before scission" and "immediately after scission" configurations lead to higher emission probabilities.}
\end{figure}

To see the effect of a larger difference between the "just before" and "immediately after" scission configurations, we have plotted in Fig.~\ref{fig:wfs6} the same states as in Fig.~\ref{fig:wfs1} for the transition $0.975 \rightarrow 1.010$. As expected, the values $P_{em}^i$ are larger (by a factor from 4 to 10) since the coefficients $|a_{if}|^2$ are now distributed over more final states.

\subsection{Scission neutron multiplicities}

Summing the partial neutron emission probabilities $P_{em}^i$ of all initially bound-states weighted by their occupation probabilities $v_i^2$, Eq.~\ref{eq:Nn}, we have obtained the total scission-neutron multiplicities $N_n$. The values obtained with several computational assumptions are found in Tables~\ref{tab:0MeV} and~\ref{tab:2MeV}.

For the dependence of $v_i^2$ on the particular state $i$, we have considered the neutrons 
\begin{enumerate}
\item independent:
\begin{eqnarray}
v_i^2 = \left\{ \begin{array}{l} 1 \mbox{  for i$\leq$144/2,} \\ 0 \mbox{ otherwise} \end{array} \right.
\end{eqnarray}
\item pairing-correlated:
\begin{eqnarray}
v_i^2=\frac{1}{2}\left[ 1-\frac{e_i-\lambda}{\sqrt{\left( e_i-\lambda \right)^2+\Delta^2}} \right]
\end{eqnarray}
\end{enumerate}

The pairing gap $\Delta$ and the Fermi energy $\lambda$ are the solutions of the two coupled-nonlinear BCS equations~\cite{Brack:1972a}. The closely related smooth gap parameter $\tilde{\Delta}\simeq 13.3/\sqrt{A}$ MeV was taken from the systematics~\cite{Moller:1992a} of odd-even mass differences in nuclei throughout the periodic table.

Tables~\ref{tab:0MeV} and~\ref{tab:2MeV} correspond to different separation lines between bound and unbound states in Eq.~\ref{eq:Pem}: 0 and 2 MeV respectively. The first value means that the centrifugal barrier is null. This is adequate when considering $\Omega=1/2$ states, since their $l$-distribution is highest at $l=0$. However, states with higher $\Omega$-values are influenced by the 2D centrifugal barrier expressed by the term $\left( \Omega\pm 1/2\right)^2/\rho^2$ present in the hamiltonian. Eigenstates with energies lower than the top of the centrifugal barrier are quasi-bound states, and could therefore be considered as still present in the fragments, their reabsorption by the fragments being large. The 2 MeV results should be viewed as an indication of the sensitivity of the results on this cut-off energy parameter. This sensitivity is rather important since the total emission probabilities are roughly reduced by half.

Due to the uneven distribution of states with given $\Omega$, the variation of $N_n$ with the energy cut-off is not monotonic. This explains why in the last three columns of Table~\ref{tab:0MeV}, $N_n$ is higher for 5/2 than for 3/2 states, contrary to the general behaviour of $N_n$ which is to decrease with $\Omega$.

Analyzing these tables, one can also notice that more than 93\% of the scission neutrons are emitted from 1/2, 3/2 or 5/2 states, more than 55\% are emitted from 1/2 states and more than 33\% from 1/2+ states. Although for most $\Omega$ values, $N_n$ is higher for positive than for negative parity states, $\Omega$=3/2 is an exception. For those states the wave functions with negative parity appear to have a maximum in the neck region. This is possible since what counts for the behaviour is the parity of the functions $d(\rho,z)$ and $u(\rho,z)$ and not of the total wave function $\Psi$ (see Eq.~\ref{eq:psi}). For instance,
\begin{eqnarray}
\pi \left( \psi_{3/2} \right) = \pi (u)\times(-)^1 \times \pi(d)\times(-)^2
\end{eqnarray}
is negative if $u(\rho,z)$ is symmetric with respect to the $\rho$-axis. It turns out that $u$ is the dominant component, i.e., $\int{u^2}>\int{d^2}$.

Tables~\ref{tab:0MeV} and~\ref{tab:2MeV} contain also a comparison between the two sets of $\left( \epsilon_i \rightarrow \epsilon_f\right)$ values considered. As expected, the calculated scission-neutron multiplicity increases with the difference ($\epsilon_f-\epsilon_i$). We found a factor of about 5 for 0-MeV cut-off and of about 6 for 2-MeV cut-off. The sensitivity of our results on the choice of $\epsilon_i$ and $\epsilon_f$ values is therefore very high, while the difference in neck radii between the two configurations is not so large (0.4 fm).

\subsection{Distribution of the emission points}

To obtain a detailed picture of scission neutron emission and to follow their motion thereafter one needs to know the distribution of the emission points.

\begin{figure}[ht]
\centerline{\includegraphics[height=0.85\textheight]{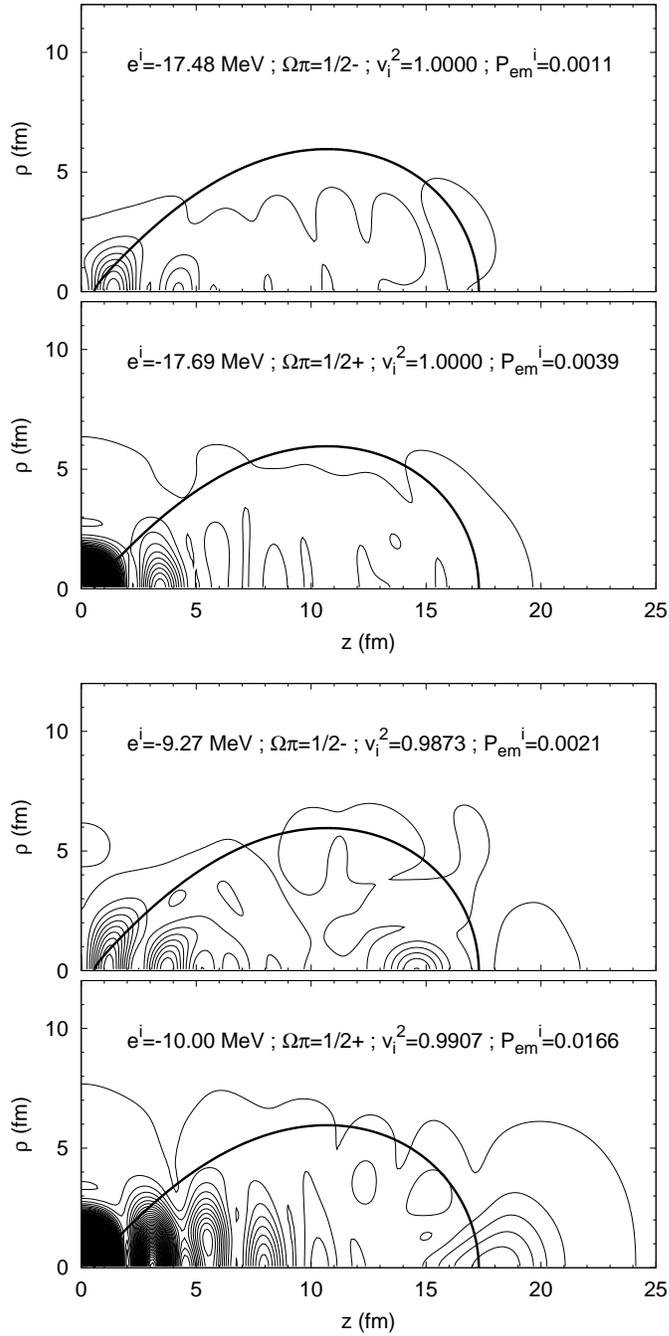}}
\caption{Emission distribution points corresponding to the initial wave functions represented in Fig.~\ref{fig:wfs1}. The solid line represents the equipotential $V_0/2$ for $\epsilon_f$=1.001.}
\label{fig:wfs1-emitted}
\end{figure}

\begin{figure}[ht]
\centerline{\includegraphics[width=0.85\textwidth]{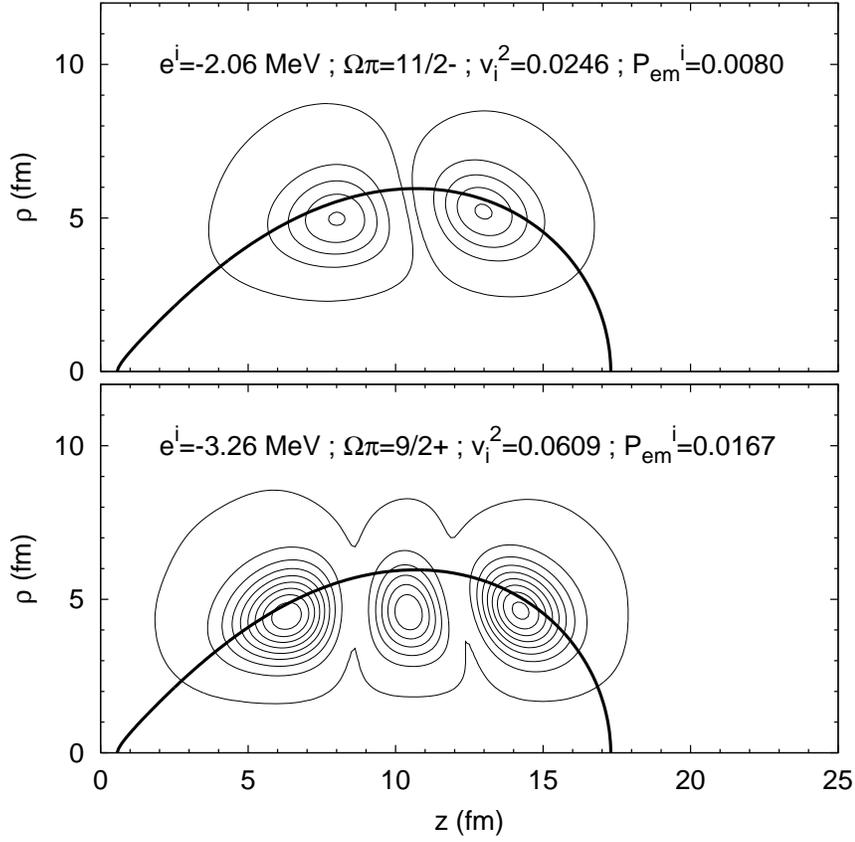}}
\caption{Emission distribution points corresponding to the initial wave functions represented in Fig.~\ref{fig:wfs5}. The emitted part of a state with high-$\Omega$ value corresponds almost perfectly to the same-$\Omega$ value state lying just above in energy. The number of nodes is simply incremented by one unit.}
\label{fig:wfs5-emitted}
\end{figure}

\begin{figure}[ht]
\centerline{\includegraphics[height=0.85\textheight]{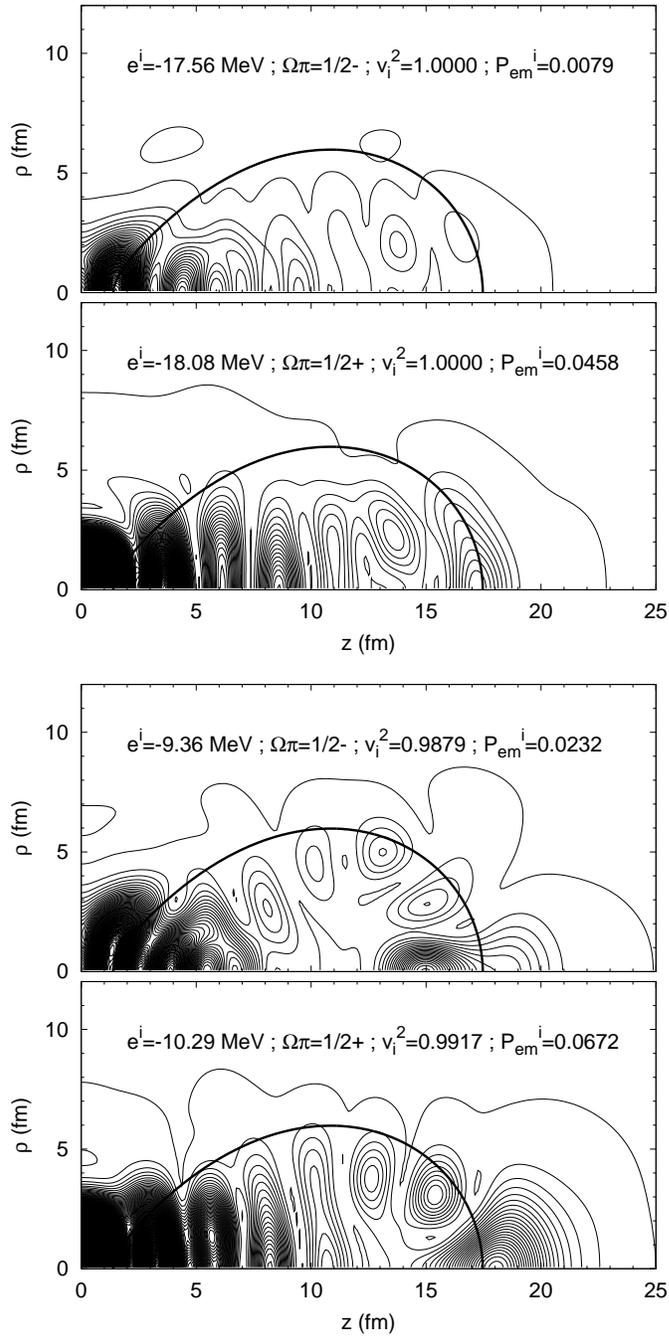}}
\caption{Emission distribution points corresponding to the initial wave functions represented in Fig.~\ref{fig:wfs6}, for the configuration $(0.975,1.010)$.}
\label{fig:wfs6-emitted}
\end{figure}

For this purpose, we have plotted in Figs.~\ref{fig:wfs1-emitted},~\ref{fig:wfs5-emitted} and~\ref{fig:wfs6-emitted} the square moduli of the emitted part of the neutron wave functions (Eq.~\ref{eq:psi_em_i}) for the same states as in Figs.~\ref{fig:wfs1},~\ref{fig:wfs5} and~\ref{fig:wfs6}. Since they represent the high-energy tails of the wave packets in the final potential well, they resemble the corresponding initial wave functions but they have more nodes (more structures in the $(\rho,z)$-plane). This is particularly clear for the highest $\Omega$ values: 9/2 and 11/2 (Fig.~\ref{fig:wfs5-emitted}). Because of the low level density of these states, the emitted part coincides with the next state in the spectrum, i.e., it has one more node than the initial state.

Those wave functions that are initially present in the neck region will have higher emission probabilities. As predicted by Halpern~\cite{Halpern:1965a}, the wave function of the emitted neutron mainly contains the part of the initial wave function that was located in the neck.

\begin{figure}[ht]
\centerline{\includegraphics[width=0.75\textwidth]{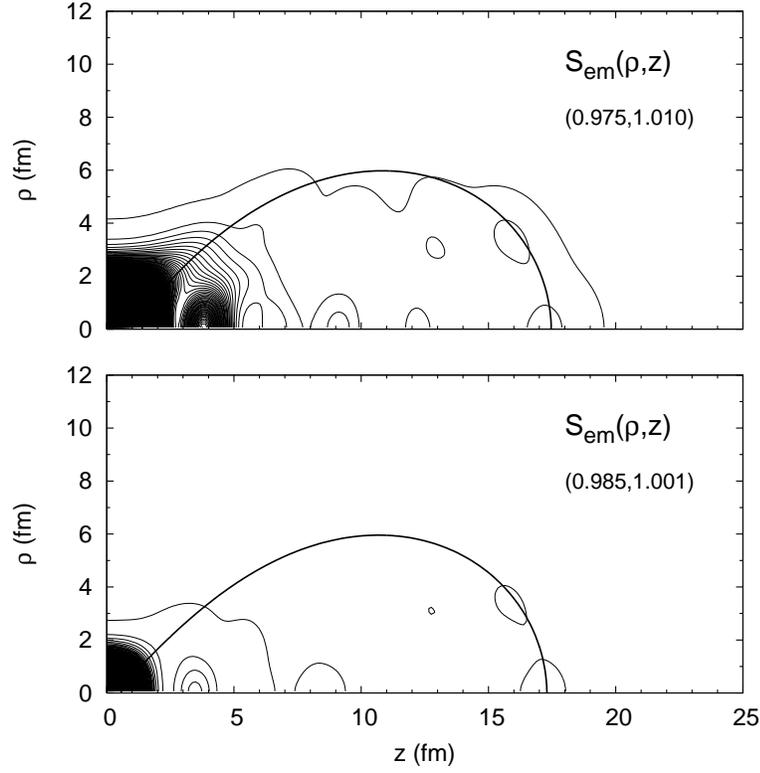}}
\caption{\label{fig:total-emission} Spatial distribution of the emission points S$_{em}(\rho,z)$ for the (0.985,1.001) [bottom] and (0.975,1.010) [top] scission configurations.}
\end{figure}

\begin{figure}[ht]
\centerline{\includegraphics[width=0.75\textwidth]{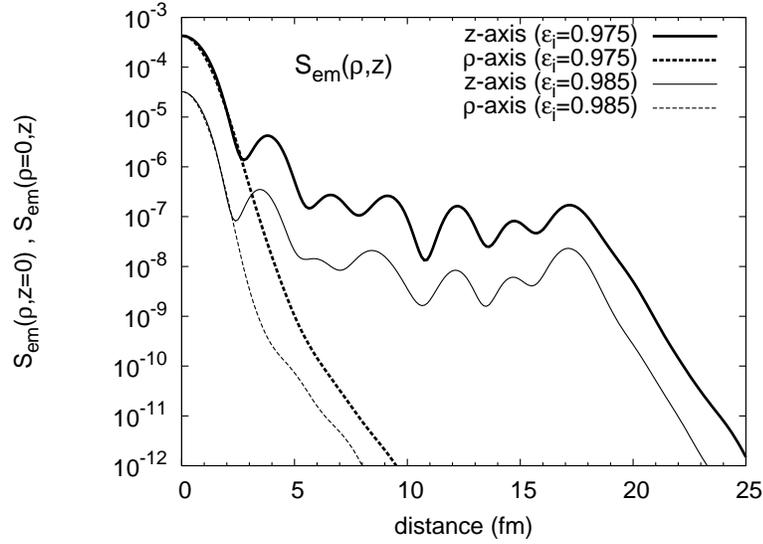}}
\caption{\label{fig:total-emission-1d} Variations of S$_{em}(\rho,z)$ along the z- and $\rho$-axes, for both scission configurations studied.}
\end{figure}

Summing over all initially-bound states like in Eq.~(\ref{eq:total-emission}), we have obtained the total distribution of the emission points (Fig.~\ref{fig:total-emission}) for the two sets of 'just before' and 'immediately after' scission configurations. Apart from the absolute values, the two distributions are similar. This can be better seen if we look only along the $\rho$ and $z$ axes as in Fig.~\ref{fig:total-emission-1d}. The large majority of the scission neutrons are emitted in the region between the fragments. To escape the nuclear re-absorption, they have to move perpendicular to the fission axis. With much lower probability (three orders of magnitude) unbound neutrons are however present also inside the fragments and at the poles of the fissioning  system. This latter component is either re-absorbed or emitted in the direction of the fission axis (polar emission).

\subsection{Primary fission fragments excitation energies}

A sudden transition at scission not only produces neutrons but also leaves the primary fragments in an excited state. This amount of excitation energy is important to know since it can be used to emit neutrons during and after the acceleration of the fragments. The later are called 'prompt' neutrons and have been extensively studied both theoretically and experimentally (for a review, see~\cite{Gonnenwein:2004a} and references therein).

\begin{figure}[ht]
\centerline{\includegraphics[width=0.75\textwidth]{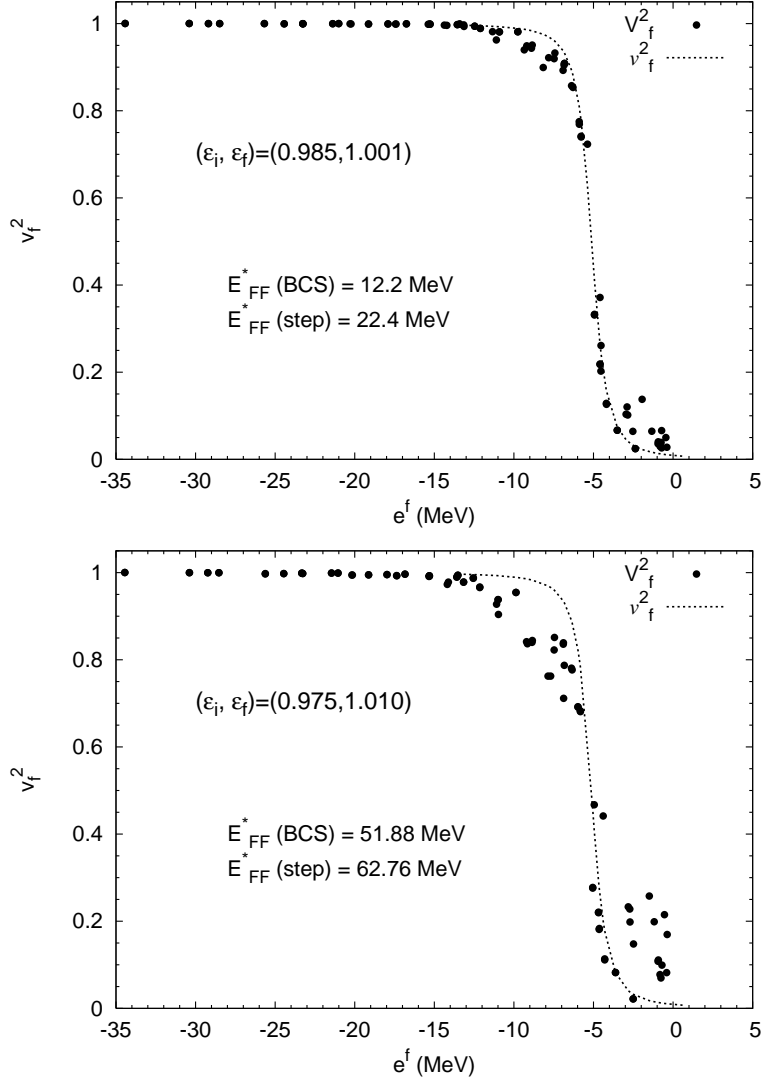}}
\caption{\label{fig:exproba} Occupation probabilities of single-neutron levels "immediately-after-scission" for the cases $(0.985,1.001)$ (top) and $(0.975,1.010)$ (bottom).}
\end{figure}

Figure~\ref{fig:exproba} shows the probabilities to occupy bound-states in the final configuration according to Eq.~(\ref{eq:exproba}). Comparing with the same probabilities in the BCS ground-state, one can see how states above the Fermi level have been populated at the expense of states below. As for the emission probabilities (Figs.~\ref{fig:Pem.985} and Fig.~\ref{fig:Pem.975}), the occupation probabilities strongly depend on the structure of the individual states involved. These structure effects are more pronounced in the lower part which corresponds to the larger difference of the deformations between which the sudden transition occurs (0.975$\rightarrow$1.010).

Using these partial occupation probabilities $V_f^2$ in Eq.~(\ref{eq:exenergy}), we have calculated the total excitation energy of the fragments immediately after scission. We found 12.2 MeV for the shorter jump and 51.88 MeV for the longer jump, i.e., roughly the same factor 4-5 as for multiplicities.

\section{Summary and conclusions}

We have presented a model calculation of quantities that characterize the nuclear configuration at scission: excitation energy of primary fission fragments, scission neutron multiplicity and initial distribution of their emission points. The approach is relatively simple and the results are easy to interpret microscopically in terms of the quantum numbers of the states involved. 

The results turned out to be highly sensitive to the value of the minimum neck radius assumed (when the neck ruptures). A precise measurement of the scission neutron multiplicity could therefore determine this important dynamical parameter. On the other hand, if the minimum neck radius is 1.5 fm (a reasonable value) 15\% of the total neutrons emitted during the symmetric fission of U-236 are scission neutrons. They are therefore non negligible and they should be explicitly taken into account in any fission data analysis. 

Our model also provided an estimate of the excitation energy of the primary fragments (12 MeV), a deciding factor for the emission of prompt neutrons, unavailable from other sources.

Due to the sudden approximation used, the values obtained represent upper limits and this fact has to be remembered when comparing with experimental data.

\section*{Acknowledgments}

One of the authors (P.T.) would like to thank Dr. Ludovic Bonneau in the Nuclear Physics Group (T-16) at Los Alamos National Laboratory for his very useful and interesting comments during the development of this work.

\newpage

\newpage

\begin{center}
\begin{table}\caption{Scission neutron multiplicities with "0-MeV" cut-off.}
 \label{tab:0MeV}
 \begin{tabular}{|c|c|c|c|c|}
\hline
$\Omega \pi$ & \multicolumn{2}{c}{$N_n$ (0.975 $\rightarrow$ 1.010)} & \multicolumn{2}{|c|}{$N_n$ (0.985 $\rightarrow$ 1.001)} \\
\cline{2-5}
& BCS & Step Function & BCS & Step Function \\
\hline
1/2 -/+ & 0.371/0.567 & 0.450/0.558 & 0.062/0.144 & 0.097/0.145 \\
3/2 -/+ & 0.214/0.082 & 0.155/0.069 & 0.028/0.013 & 0.017/0.010 \\
5/2 -/+ & 0.188/0.202 & 0.230/0.241 & 0.042/0.043 & 0.051/0.052 \\
7/2 -/+ & 0.047/0.047 & 0.007/0.007 & 0.011/0.011 & 0.001/0.001 \\
9/2 -/+ & 0.006/0.006 & 0.002/0.002 & 0.001/0.001 & 0.000/0.000 \\
11/2 -/+ & 0.001/0.001 & 0.000/0.000 & 0.000/0.000 & 0.000/0.000 \\
\hline
Total -/+ & 0.828/0.904 & 0.845/0.878 & 0.144/0.213 & 0.168/0.209 \\
\hline
Total & 1.733 & 1.723 & 0.358 & 0.377 \\
\hline
\end{tabular}
\end{table}
\end{center}

\bigskip

\begin{center}
\begin{table}\caption{Scission neutron multiplicities with "2-MeV" cut-off.}
\label{tab:2MeV}
\begin{tabular}{|c|c|c|c|c|}
\hline
$\Omega \pi$ & \multicolumn{2}{c}{$N_n$ (0.975 $\rightarrow$ 1.010)} & \multicolumn{2}{|c|}{$N_n$ (0.985 $\rightarrow$ 1.001)} \\
\cline{2-5}
& BCS & Step Function & BCS & Step Function \\
\hline
1/2 -/+ & 0.259/0.400 & 0.306/0.381 & 0.032/0.089 & 0.045/0.085 \\
3/2 -/+ & 0.187/0.072 & 0.139/0.062 & 0.021/0.012 & 0.014/0.009 \\
5/2 -/+ & 0.027/0.032 & 0.030/0.033 & 0.004/0.004 & 0.005/0.005 \\
7/2 -/+ & 0.012/0.012 & 0.002/0.002 & 0.003/0.003 & 0.001/0.001 \\
9/2 -/+ & 0.006/0.006 & 0.002/0.002 & 0.001/0.001 & 0.000/0.000 \\
11/2 -/+ & 0.001/0.001 & 0.000/0.000 & 0.000/0.000 & 0.000/0.000 \\
\hline
Total -/+ & 0.493/0.523 & 0.480/0.481 & 0.062/0.110 & 0.065/0.099 \\
\hline
Total & 1.016 & 0.960 & 0.172 & 0.164 \\
\hline
\end{tabular}
\end{table}
\end{center}

\newpage

\newpage

\newpage

\newpage

\newpage

\newpage

\newpage

\newpage

\newpage

\newpage

\newpage




\newpage

\newpage

\newpage

\end{document}